\documentclass[apl,reprint,amsmath,amssymb]{revtex4-1}
\usepackage[caption=false]{subfig}
\usepackage{graphicx}% Include figure files
\usepackage{dcolumn}% Align table columns on decimal point
\usepackage{bm}% bold math
\usepackage{longtable}
\usepackage{xcolor}
\begin{document}

\preprint{APS}

\title{Fractional Fresnel coefficients for optical absorption in femtosecond laser-induced rough metal surfaces}
	%Laser Absorption at Rough Metal Surfaces with Fractal Features\\OR\\
	%Modeling the Effects of Laser-Induced Rough Fractal Surface Structures on %Optical Absorptance of Metals
	%\\ Effect of Fractal Features of Rough Surfaces on Optical Absorption of Metals\\OR
	% Force line breaks with \\
%\thanks{title to be modified later}%

 \author{Muhammad Zubair}%
 \email{muhammad.zubair@itu.edu.pk}
 \affiliation{%
 	Electrical Engineering Department, Information Technology University of the Punjab, Lahore, Pakistan 54600\\SUTD-MIT International Design Center, Singapore University of Technology and Design, Singapore 487372
 %\textbackslash\textbackslash\\
 }%

 %\thanks{\\M. Zubair is now with the Electrical Engineering Department of Information Technology University of the Punjab, Lahore, Pakistan. }%
 \author{Yee Sin Ang}%
 \author{Kelvin J. A. Ooi}%
\author{L. K. Ang}%
 \email{ricky$_$ang@sutd.edu.sg}

\affiliation{%
 Science and Math \& SUTD-MIT International Design Center, Singapore University of Technology and Design, Singapore 487372%\textbackslash\textbackslash\\
}%

%\affiliation{
%\affiliation{
%\affiliation{%


\begin{abstract}
The surface morphology of metal influences its optical absorptivity. 
Recent experiments have demonstrated that the femtosecond laser induced surface structures on metals could be dynamically controlled by the fluence of laser and the number of pulses.
In this paper, we formulate an analytical model to calculate the optical absorption of a rough metallic surface by modeling the roughness as a fractal slab.
For a given experimental image of the surface roughness, we characterize the roughness with a fractal parameter by using box-counting method. 
With this parameter as input, we calculate the absorption of 800 nm laser pulse impinging on gold, copper and platinum, and the calculated results show excellent agreements.
In terms of physics, our model can be viewed as a fractional version of the Fresnel coefficients, and it will be useful for designing suitable surface structures to tune the light absorption on metals from purely reflective to highly absorptive based on different applications.
\end{abstract}

\maketitle

%\tableofcontents

\section{\label{sec:Introduction} Introduction}
Metal is a highly reflective material, however it is important to increase its optical absorption for some specific applications ~\cite{gamaly2011femtosecond, vorobyev2013direct}. 
It was demonstrated that the optical absorption of metals can be significantly enhanced by using femtosecond (fs) laser induced surface structures ~\cite{vorobyev2005enhanced, vorobyev2009enhanced}.
As an example, for gold, the absorption can be increased from its intrinsic value of a few percent to nearly perfect absorption due to different laser-induced surface roughness ~\cite{vorobyev2005enhanced}.
The temporal and spatial evolution of this type of fs laser-induced roughness on metals by utilizing a time-resolved optical imaging technique has been obtained ~\cite{fang2017direct}.
Under multiple pulses condition, nano-scale and micro-scale structures are respectively, formed at lower and higher laser fluence. 
Compared to other techniques, the fs laser pulse processing technique has received attention over the past decades owing to its ability to apply on multi-scale non-planar surfaces of metals~\cite{sugioka2014ultrafast}.

Depending on the size of roughness, which can be either smaller or larger than laser wavelength ($\lambda$), the physics of the enhanced absorption of laser can be due to different mechanisms. 
For roughness smaller than $\lambda$, the light absorption can be due to the antireflection effect of random sub-wavelength surface textures in terms of a graded refractive index at the air-metal interface~\cite{ghmari2004influence}. 
For roughness greater than $\lambda$, the enhancement can be due to multiple reflections within the large surface structures~\cite{ang1997analysis}.  
The optical properties of a geometrical modified nano-materials is also different from its bulk counterpart~\cite{gavrilenko2011optics}. 

   % \onecolumngrid

     \begin{figure}[!hb]
     	\centering
     	\includegraphics[width=0.48\textwidth]{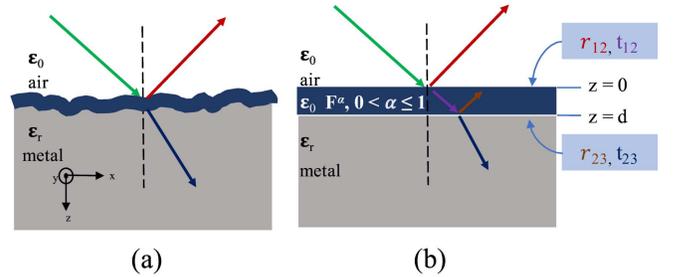}
     	\caption{(a) Diagram of laser absorption on a rough metal surface. (b) Effective model, where the air-metal interface is treated as a slab of thickness $d$ in a fractional-dimension-space ($F^{\alpha}$) and the value of $\alpha < 1$ measures the effect of roughness. The value of $d$ is determined by having the calculation at $\alpha$ = 1 (zero roughness) equal to the intrinsic absorption value or experimental measurement.}
     	\label{fig:geometry}
     \end{figure}
 %        \twocolumngrid

It is computationally expensive to apply full-wave numerical simulation to calculate the absorption of light over a wide range of roughness scales.
In the existing analytical or semi-analytical models~\cite{ang1997analysis,bergstrom2008absorption}, the characterization of roughness and surface morphology is approximated by various profilometric measurements and mathematical models, which are expressed as a function of various surface roughness parameters, such as depth ($h$), width ($w$) and its ratio of $h/w$.
Given the wide range of roughness (from nano- to micro- and macro-scale), the prior models may not be sufficient to describe the complexity of a rough surface.
Our analysis on several SEM images of such rough surfaces has shown that the surface roughness can be better characterized as fractals~\cite{mandelbrot1983fractal} (see Supplementary Information). 
The fractal description is insensitive to structural scale and provides a useful parameter to characterize the surface morphology, which is termed as the fractal dimension~\cite{falconer2004fractal}. 
The fractal-based model is also independent of the resolution of the characterization~\cite{risovic2009correlation}. 
It has been shown statistically that the fractal dimension is one of the most relevant parameters in surface morphology characterization~\cite{risovic2008inferring} that can also be correlated with various other surface roughness parameters~\cite{russ2001fractal}. 

In this paper, we develop an analytical fractional model to calculate the absorption of a laser pulse impinging on a rough metal, where the effect of roughness is treated by using a fractal parameter called $\alpha$ ($\leq 1$). 
By solving the corresponding Maxwell equations with appropriate boundary conditions in a fractional model (see below), we obtain an analytical form of the fractional Fresnel coefficients.
By analyzing the SEM images of laser-induced rough metal surface structures from various experiments of different metals ~\cite{vorobyev2007change,vorobyev2005enhanced,vorobyev2007effects}, we mathematically determine the values of $\alpha$, and use them to calculate the optical absorption, which agree very well with experimental measurement.
At $\alpha$ = 1, the model will reproduce the absorption for a perfect surface with zero roughness.
It is important to note that our model provide (for the first time) the analytical solutions to relate directly the laser absorption to the roughness of a metal surface.
The proposed model is useful in the design of different surface structures on any metal surfaces so one can tune the optical absorption on a metallic surface in many applications such as photonics, plasmonics, optoelectronics, optofluidics, stealth technology, airborne devices, solar energy absorbers, and thermo-photovoltaics (see~\cite{vorobyev2013direct,sugioka2014ultrafast,fang2017direct}; and reference therein).

%Recently, femtosecond (fs) laser-induced surface structuring has emerged as a novel and versatile technology 
%Due to its ability to process non-planar surfaces of nearly all types of solid materials (including metals, semiconductors, glasses, and polymers) in a maskless single-step process at high speed under normal conditions, this technique has received a great amount of research attention over the past decade~\cite{sugioka2014ultrafast}. 

\section{Fractional Model of Optical Absorption} 
Fractional-dimensional approach has attracted widespread attention in recent years motivated by its fundamental importance and possible practical applications in electromagnetic modeling of complex, anisotropic, heterogeneous, disordered or fractal media~\cite{zubair2012electromagnetic,tarasov2014anisotropic}. This approach has been applied in various areas of physics and engineering
including the quantum field
theory~\cite{stillinger1977axiomatic,palmer2004equations},
general relativity~\cite{sadallah2009solution},
thermodynamics~\cite{tarasov2016heat},
mechanics~\cite{ostoja2014fractal},
hydrodynamics~\cite{balankin2012map},
electrodynamics~\cite{tarasov2015electromagnetic, zubair2012electromagnetic,mughal2011fractional,naqvi2016cylindrical, zubair2011exact, asad2012electromagnetic, asad2012reflection,zubair2011exact2,zubair2011differential,zubair2011electromagnetic,zubair2010wave}, and
fractional charge transport~\cite{zubair2016fractional,zubair2017fractional,zubair2018thickness} to name a few.

The proposed model is based on the formulation of Maxwell equations in fractional-dimensional spaces~(see~\cite{zubair2012electromagnetic} and references therein) to calculate the effective Fresnel coefficients for a generalized rough surface as shown in Fig. (\ref{fig:geometry}). The rough metal surface is modeled as a slab of thickness $d$ from $z=0$ to $z=d$, where we have air and metal, respectively, in region $z < 0$ (above) and $z > d$ (below). 
For an interface at $z=d$, we define two fractional-dimensional (FD) regions, with fractal dimensions $0<D_{a}\leq3$ and $0<D_{b}\leq3$.
The region of $0 < z<d$, it has a dimension of $D_{a}$ for a medium with parameters $\epsilon_{a}$, and $\mu_{a}$.
Similarly, for $z>d$, we have dimension $D_{b}$ for a medium with parameters $\epsilon_{b}$ and $\mu_{b}$. For general formulation, we keep the expressions of $\epsilon_{a}$, $\mu_{a}$, $\epsilon_{b}$, and $\mu_{b}$, instead of spelling out the medium as air or metal specifically.
\textcolor{red}{In this simplistic fractional model, it is considered that the surface roughness alters the effective EM field in the direction normal to the interface and hence only the normal coordinate is fractionalized to achieve an equivalent geometry shown in Fig. (\ref{fig:geometry}).  It is to be noted that this model does not provide exact field distribution at the rough surface locally, however, an approximate value of variation in absorption can be simply achieved in an effective manner.} Hence, under the assumption that both regions are fractional dimensions only in the direction normal to the interface ($z$-axis), we may write $D_{a,b}=2+\alpha_{a,b}$, where both $0<\alpha_{a}\leq 1$ and  $0<\alpha_{b}\leq 1$ are the corresponding fractal dimensions. 

Considering a linearly polarized transverse magnetic (TM) or p-polarized plane wave with an electric-field strength $E_o$ incident at $z = 0$.
The incident, reflected and transmitted electric fields are given, respectively, as
 \begin{eqnarray}
&&\textbf E_{i}=E_{0}A(\hat{a}_{x}\cos\theta_{i}-\hat{a}_{z}\sin\theta_{i})\exp(-j\beta_{a}x\sin\theta_{i}),\\
&&\textbf E_{r}=\Gamma_{ab}E_{0}B(\hat{a}_{x}\cos\theta_{r}+\hat{a}_{z}\sin\theta_{r})\exp(-j\beta_{a}x\sin\theta_{r}),\nonumber\\\\
&&\textbf E_{t}=T_{ab}E_{0}C(\hat{a}_{x}\cos\theta_{t}-\hat{a}_{z}\sin\theta_{t})\exp(-j\beta_{b}x\sin\theta_{t}),
 \label{eqn:E_fields}
 \end{eqnarray}
with,
 \begin{eqnarray}
&& A=(\beta_{a}d\cos\theta_{i})^{v_{a}} H_{v_{a}}^{(2)}(\beta_{a}d\cos\theta_{i}),\\
&& B=(\beta_{a}d\cos\theta_{r})^{v_{a}} H_{v_{a}}^{(2)}(\beta_{a}d\cos\theta_{r}),\\
&&   C=(\beta_{b}d\cos\theta_{t}) ^{v_{b}}H_{v_{b}}^{(2)}(\beta_{b}d\cos\theta_{t}).
 \label{eqn:E_fields_constant}
 \end{eqnarray}
Here, $\beta_{a,b}=\omega\sqrt{\epsilon_{a,b}\mu_{a,b}}$ is the corresponding wave number of each half space, $H_{v_{a,b}}^{(2)}(\cdot)$ is the Hankel function of type 2 with order $v_{a,b}=1-\alpha_{a,b}/2$. 
The angle of incidence, reflection and transmission are $\theta_{i}$, $\theta_{r}$ and $\theta_{t}$.
Similarly, the corresponding magnetic fields are 
  \begin{eqnarray}
 &&\textbf H_{i}=\frac{E_{0}}{\eta_{a}}D\exp(-j\beta_{a}x\sin\theta_{i})\hat{a}_{y},\\
&&\textbf H_{r}=-\Gamma_{ab}\frac{E_{0}}{\eta_{a}}E\exp(-j\beta_{a}x\sin\theta_{r})\hat{a}_{y},\\
&&\textbf H_{t}=T_{ab}\frac{E_{0}}{\eta_{b}}F\exp(-j\beta_{a}x\sin\theta_{t})\hat{a}_{y},
  \label{eqn:H_fields}
  \end{eqnarray}
  with,
  \begin{eqnarray}
  && D=(\beta_{a}d\cos\theta_{i})^{v_{ah}} H_{v_{ah}}^{(2)}(\beta_{a}d\cos\theta_{i}),\\
  && E=(\beta_{a}d\cos\theta_{r})^{v_{ah}} H_{v_{ah}}^{(2)}(\beta_{a}d\cos\theta_{r}),\\
  &&  F=(\beta_{b}d\cos\theta_{t}) ^{v_{bh}}H_{v_{bh}}^{(2)}(\beta_{b}d\cos\theta_{t}).
  \label{eqn:H_fields_constant}
  \end{eqnarray}
Here, we have $v_{ah,bh}=|v_{a,b}-1|$ and the wave impedance is $\eta_{a,b}=\sqrt{\mu_{a,b}/\epsilon_{a,b}}$. 

The transmission ($T$) and reflection ($\Gamma$) coefficients are obtained by equating the tangential components of the electric and magnetic fields at the interfaces. 
By using the Snell's law of $\beta_{a}\sin\theta_{i}=\beta_{a}\sin\theta_{r}=\beta_{b}\sin\theta_{t}$, we have
  \begin{eqnarray}
\Gamma_{ab}(\theta_{i},d)=\frac{\eta_{b}\cos\theta_{t}DC-\eta_{a}\cos\theta_{i}AF}{\eta_{a}\cos\theta_{i}BF+\eta_{b}\cos\theta_{t}CE},  \label{eqn:Fresenal_Coeff_P-pol}
\\
T_{ab}(\theta_{i},d)=\frac{\eta_{b}\cos\theta_{i}(AE+BD)}{\eta_{a}\cos\theta_{i}BF+\eta_{b}\cos\theta_{t}CE}.
  \end{eqnarray}
By carrying out the same procedure, for transverse electric (TE) polarization or s-polarization, we get
   \begin{eqnarray}
   \Gamma_{ab}(\theta_{i},d)=\frac{\eta_{b}\cos\theta_{i}DC-\eta_{a}\cos\theta_{t}AF}{\eta_{b}\cos\theta_{i}CE+\eta_{a}\cos\theta_{t}BF},\\
   T_{ab}(\theta_{i},d)=\frac{\eta_{b}\cos\theta_{i}(AE+BD)}{\eta_{b}\cos\theta_{i}CE+\eta_{a}\cos\theta_{t}BF}.
   \label{eqn:Fresenal_Coeff_s-pol}
   \end{eqnarray}

As mentioned before, the absorption at a rough metal interface is calculated effectively by using a thin slab of fractional dimension $\alpha$ and of thickness $d$ as shown in Fig. (\ref{fig:geometry}). 
The region 1 ($z<0$) is air, the region 2 ($0<z\leq d$) is a fractional slab ($0<\alpha\leq1$), and the region 3 ($z>d$) corresponds to metal characterized by complex permittivity $\epsilon_{r}$. 
The transmission and reflection coefficients for interface at $z=0$ are represented as $t_{12}$ and $r_{12}$, which can be both calculated using the expressions of $T$ and $\Gamma$ given in Eq. (\ref{eqn:Fresenal_Coeff_P-pol}) through Eq. (\ref{eqn:Fresenal_Coeff_s-pol}) by setting appropriate input parameters ($d$, $\alpha$) for a given polarization and the corresponding $\mu$ and $\epsilon$.
Similarly, we can calculate $t_{23}$, $r_{23}$, respectively, at $z = d$.

Using the calculated $t_{12}$, $r_{12}$, $t_{23}$, $r_{23}$, the overall reflectivity $R$ and absorptivity $A$ become ~\cite{born2013principles}
    \begin{eqnarray}
        R(\theta_{i})=\left|\frac{r_{12}+r_{23}}{1+r_{12}r_{23}}
        \right|^2,   \\
    A(\theta_{i})=\left|\frac{t_{12}t_{23}}{1+r_{12}r_{23}}
\right|^2.    \label{eqn:abs_coeeficient}
    \end{eqnarray}
    
 Note the value of $d$ is determined by matching the calculated absorption at $\alpha$ = 1 (zero roughness) to the theoretical value or experimental measurement for a flat surface.

\section{ Results and Discussions}

\begin{figure}[!hb]
    	\subfloat[]{%
    		\includegraphics[width=.33\textwidth]{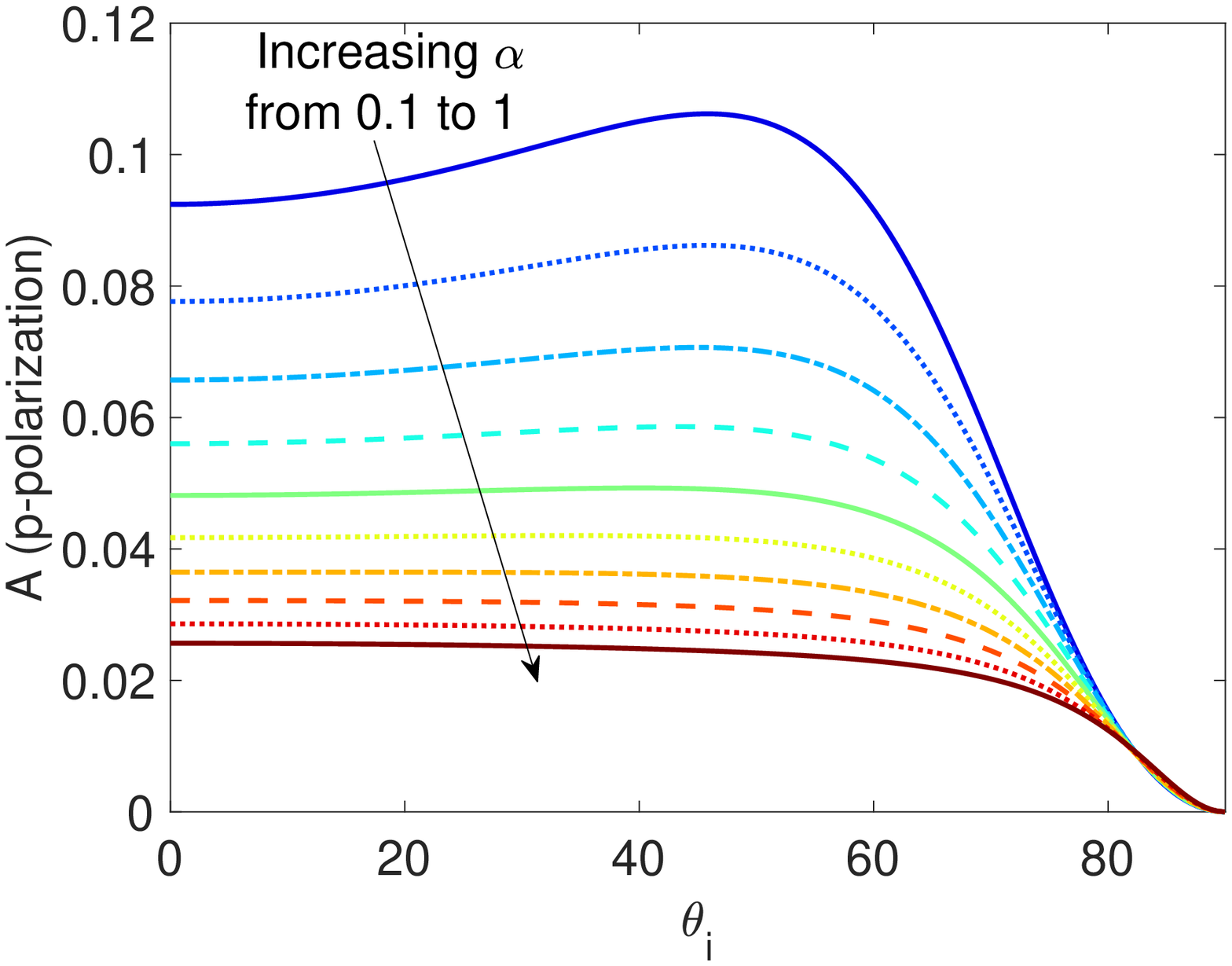}%
    	%	\label{Fig:Alpha0x4}
    	}
    		\hfill
    	\subfloat[]{%
    		\includegraphics[width=.33\textwidth]{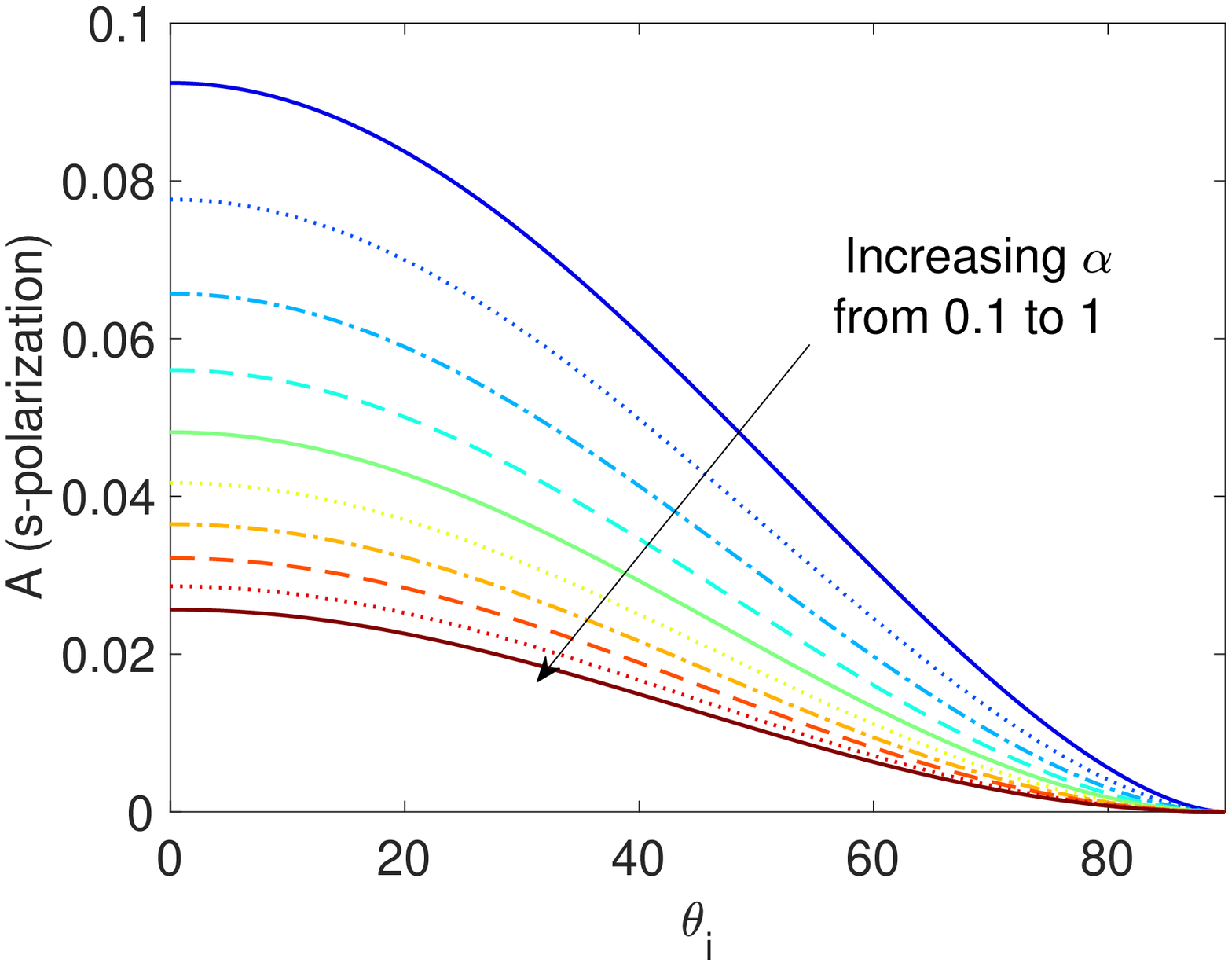}%
    		%\label{Fig:Alpha1}
    	} 
    	%\hfill
    	\caption{Absorption of laser radiation of $\lambda$=690nm on a solid aluminium target for (a) p-polarization, (b) s-polarization.} \label{Fig:Absorption_Al_Ricky}
    \end{figure}

Figure (\ref{Fig:Absorption_Al_Ricky}) shows the calculated absorptivity from Eq. (\ref{eqn:abs_coeeficient}) for s and p-polarized laser radiation of $\lambda$=690nm imping on a solid aluminium target as a function of incidence angle for different fractional parameter $\alpha$ = 0.1 to 1.
From the figure, we see the absorption increases with roughness when $\alpha$ is reduced from 1 to 0.1.
The limit of $\alpha=1$ corresponds to a perfectly smooth surface with an absorption of around 2 percent.

To compare with experimental results, we analyze the reported experimental absorption data of fs laser-induced roughness on 3 metals: copper \cite{vorobyev2007change}, gold \cite{vorobyev2005enhanced} and platinum~\cite{vorobyev2007effects}, all under 800 nm laser incidence at zero incidence.
From the experiments, the surface structures were produced by varying laser fluence ($F$) and by using different number of laser pulses ($N$). 
For comparison, we first determine the surface fractal-dimension $D$ from the published SEM images by using \textcolor{red}{box-counting method~\cite{moisy2008computing,davis2017wetting,chen2017comparison}} to determine the dimension of a fractal object ~\cite{falconer2004fractal}.

\begin{figure}[!ht]
	\subfloat[\label{fig:cu_SEM}]{%
		\includegraphics[width=.15\textwidth]{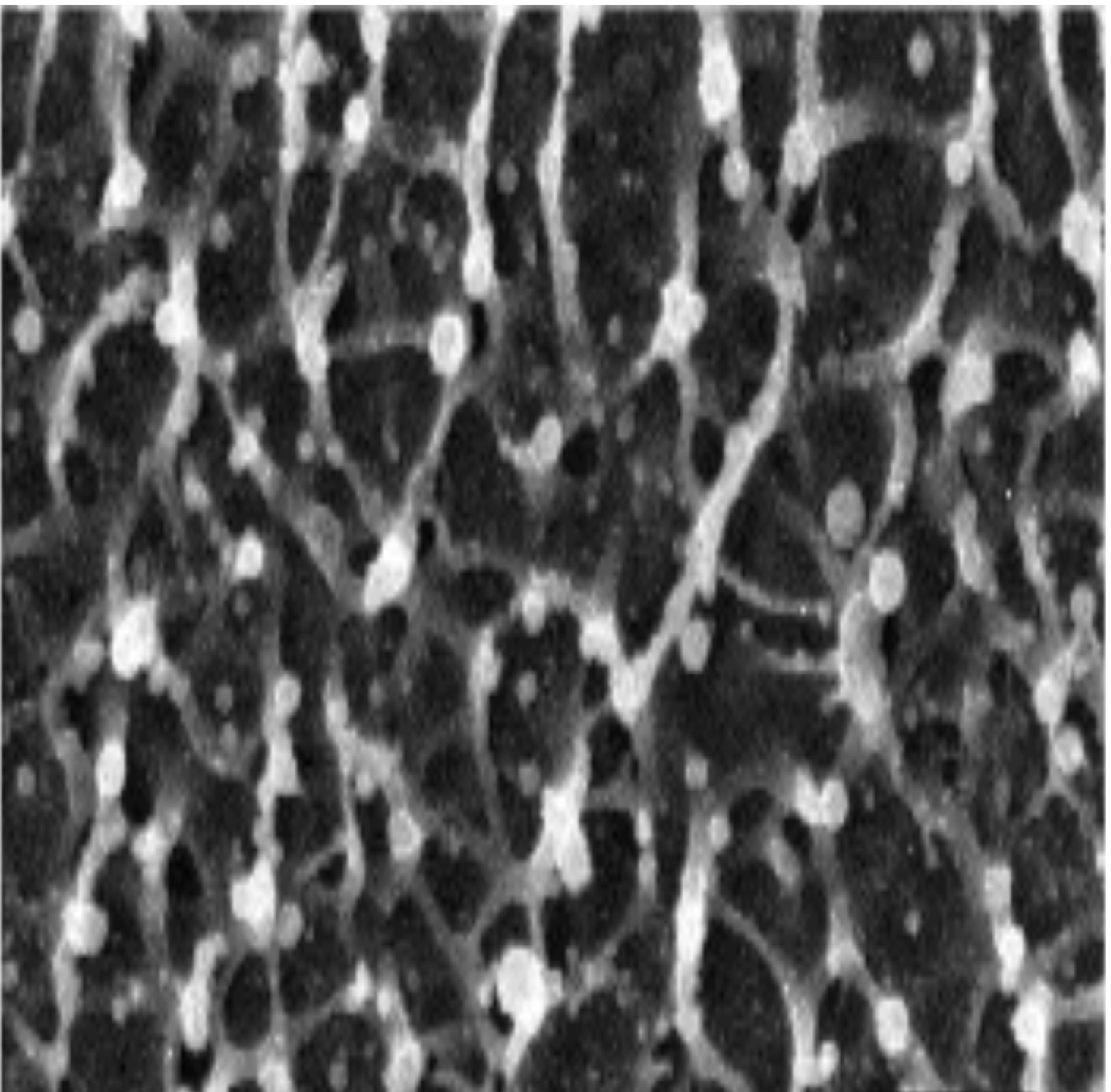}%
		%	\label{Fig:Cu}
	}
	\hfill
	\hfill
	\hfill
	\subfloat[\label{fig:cu_box}]{%
		\includegraphics[width=.18\textwidth]{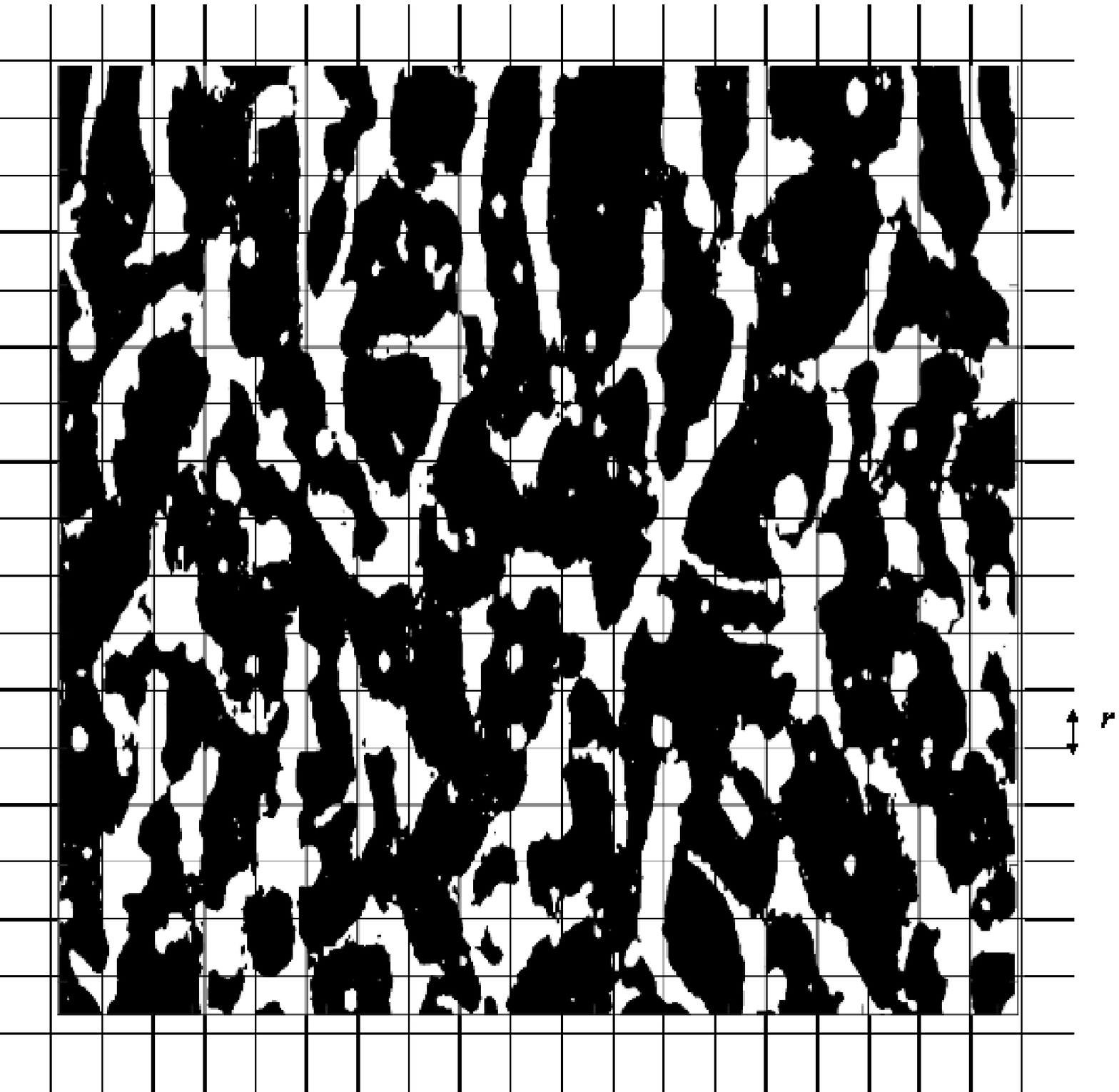}%
		%	\label{Fig:Au}
	} 
	\subfloat[\label{fig:cu_dim}]{%
		\includegraphics[width=.22\textwidth]{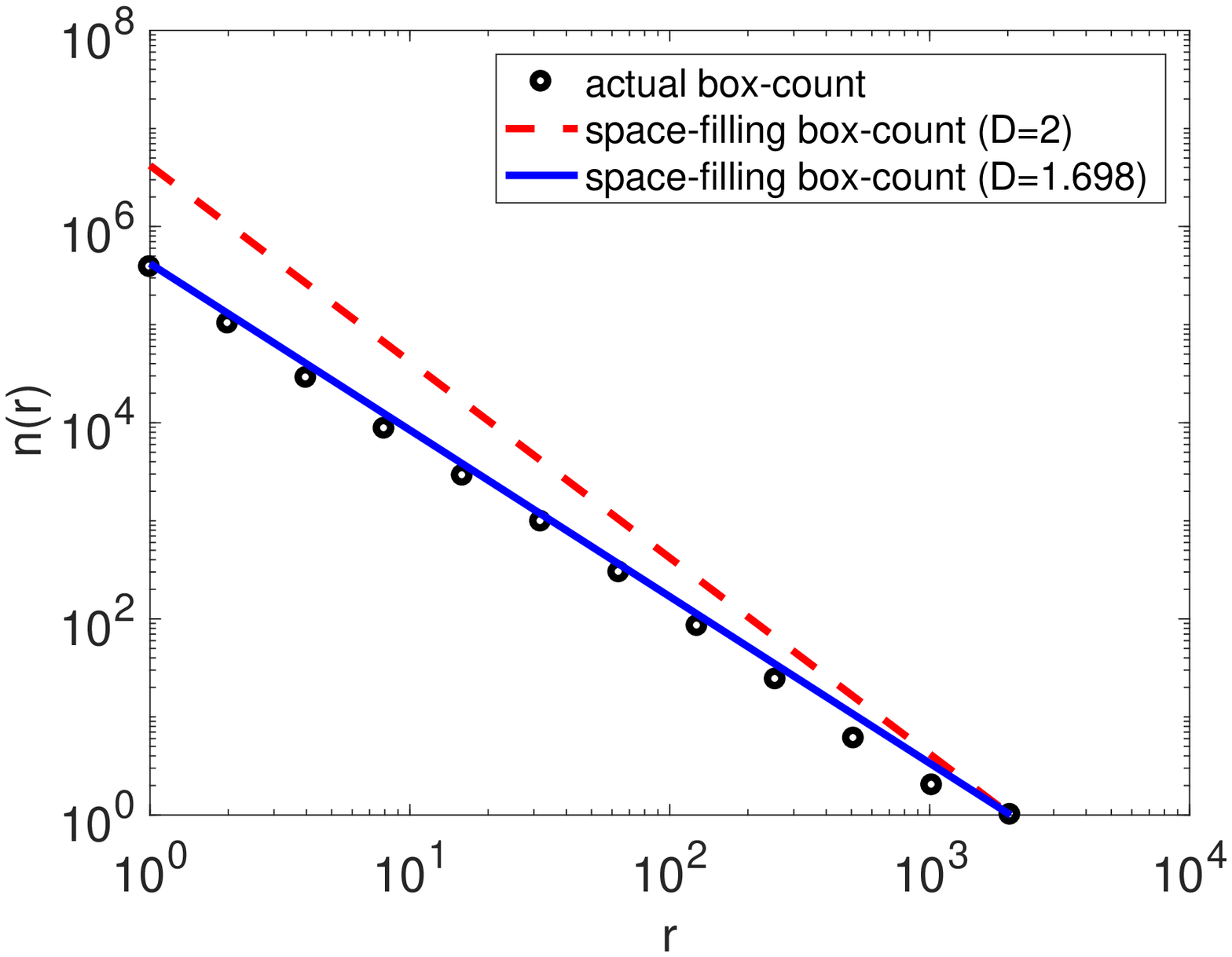}%
		%\label{Fig:Pt}
	} 
	\caption{(a)  SEM image of a rough copper surface from Ref.~\cite{vorobyev2007change} for $F = 1.52~     \mathrm{J/cm^2}$ and $N=1$ [Reproduced from IOP Open Access Journal of Physics: Conference Series] (b)  SEM image after digitization and conversion into a binary image which is broken up into a grid of boxes of varying size ($r$).(c) The number of boxes ($n$) vs the box sizes ($r$) for fractal-dimension calculation.} \label{Fig:box_counting}
\end{figure}     

In doing so, the SEM images of rough metal samples are first digitized in the form of binary images (black and white) as shown in Fig.~(\ref{fig:cu_SEM}) for a rough copper taken from~\cite{vorobyev2007change}, and it is converted then into a binary image before dividing it into a grid of boxes as shown in Fig.~(\ref{fig:cu_box}). 
The number $n$ of the boxes of size $r$ needed to cover a fractal object follows a power-law of $n(r) \propto r^{-D}$, with $D$ being the box-counting fractal dimension. 
As an example, Fig.~(\ref{fig:cu_dim}) shows that the slope of $n(r)$ versus $r$ on log-scale, which gives the fractal dimension $D$. 
For this image, the fractional $D$ is 1.698 (solid line), which is less than a flat surface of $D$ = 2 plotted in dashed lines. 

Note that in the general formulation above, we define $0 < D\leq 3$, with $\alpha = D -2$. 
However in the characterization of 2D images, the dimension is limited to $0 < D \leq 2$, so the fractional parameter $\alpha$ is adjusted accordingly to be $\alpha = D-1$.
The rough copper image of $D$ = 1.698 corresponds to $\alpha$ = 0.698, while a perfect flat surface ($D$ = 2) has $\alpha = 1$.

\begin{figure}[!h]
       	\subfloat[\label{fig:cu_exp}]{%
       		\includegraphics[width=.25\textwidth]{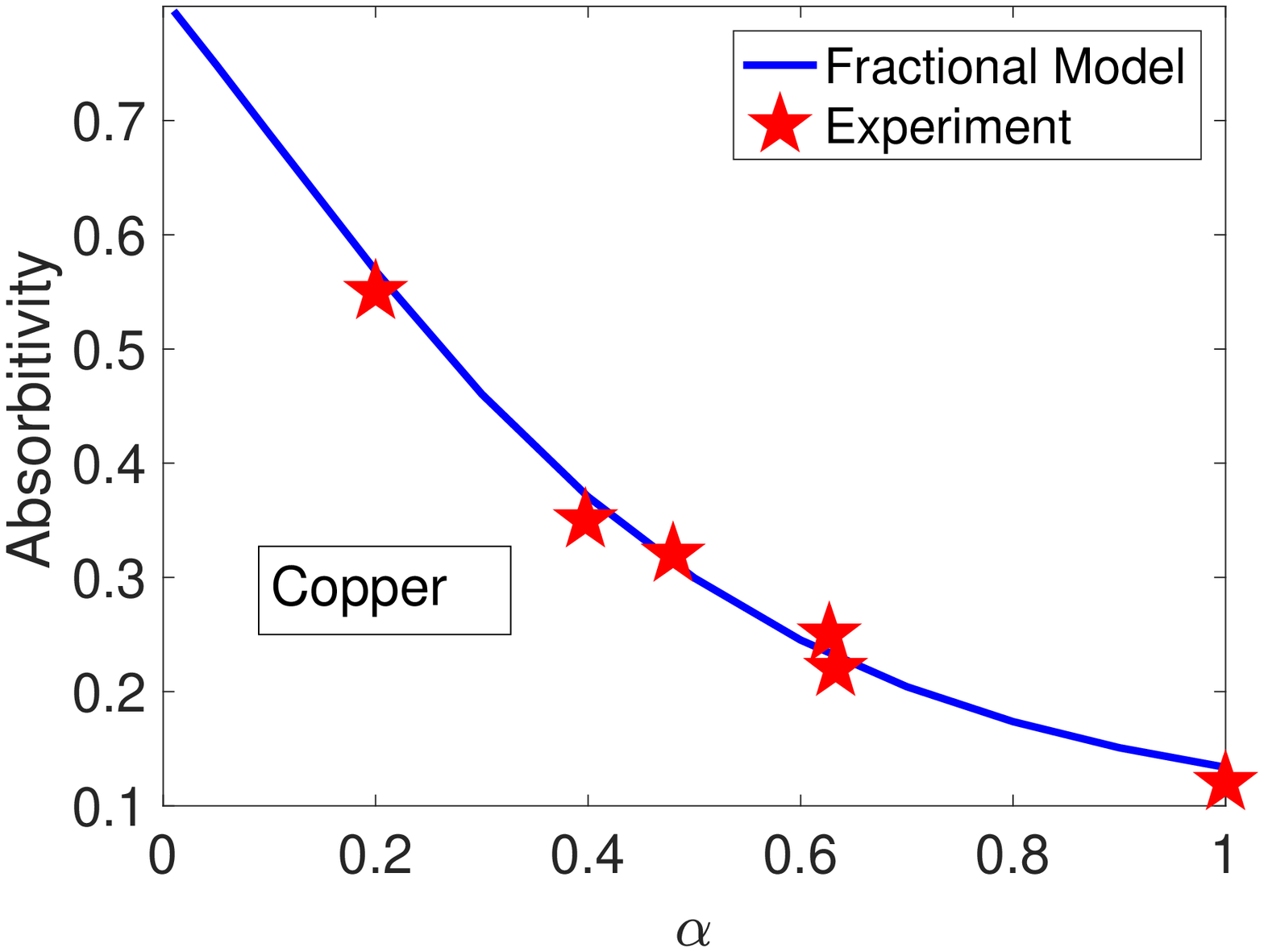}%
       		%	\label{Fig:Cu}
       	}
       	%	\hfill
       	\subfloat[\label{fig:au_exp}]{%
       		\includegraphics[width=.25\textwidth]{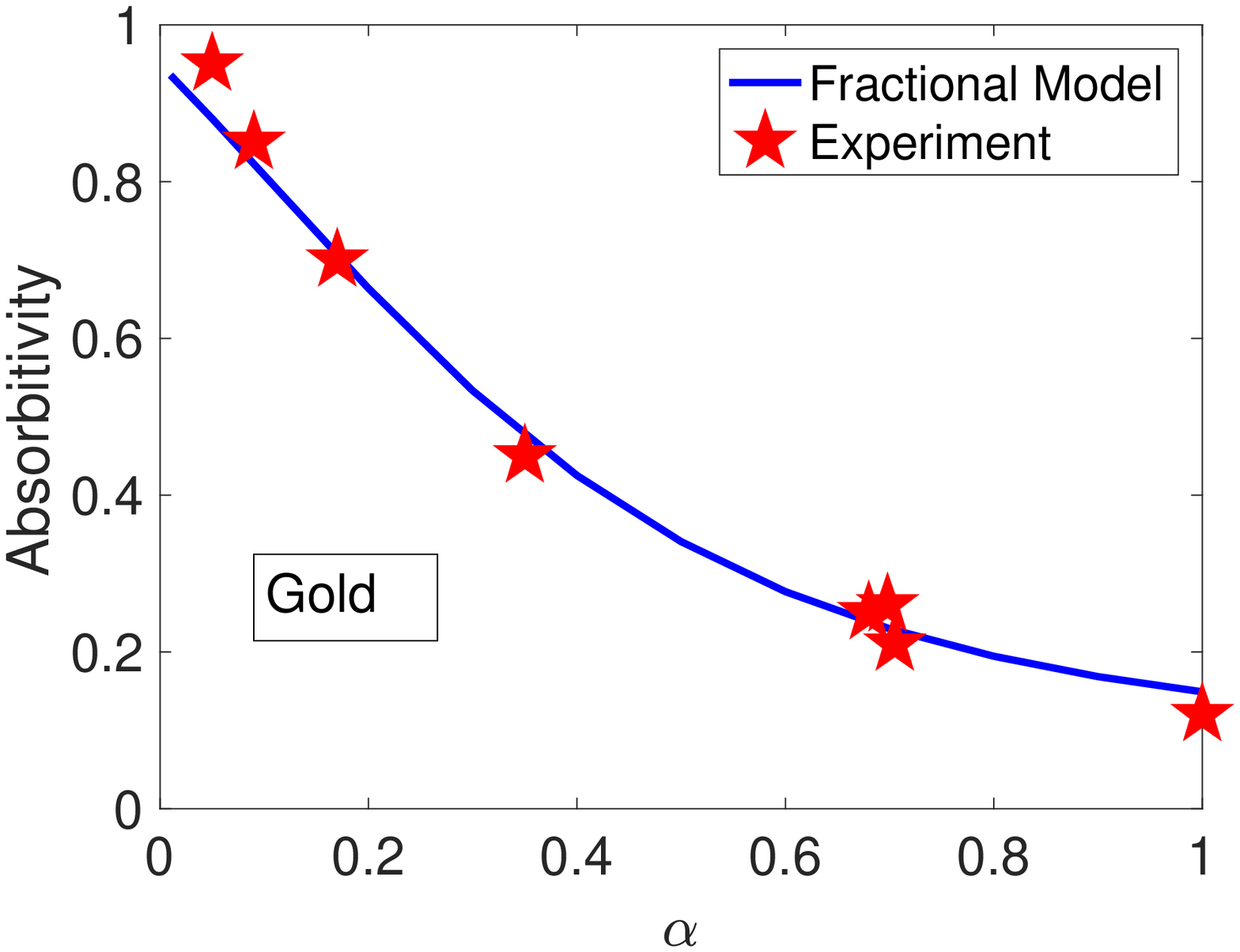}%
       		%	\label{Fig:Au}
       	} 
       	\hfill
       	\subfloat[\label{fig:pt_exp}]{%
       		\includegraphics[width=.30\textwidth]{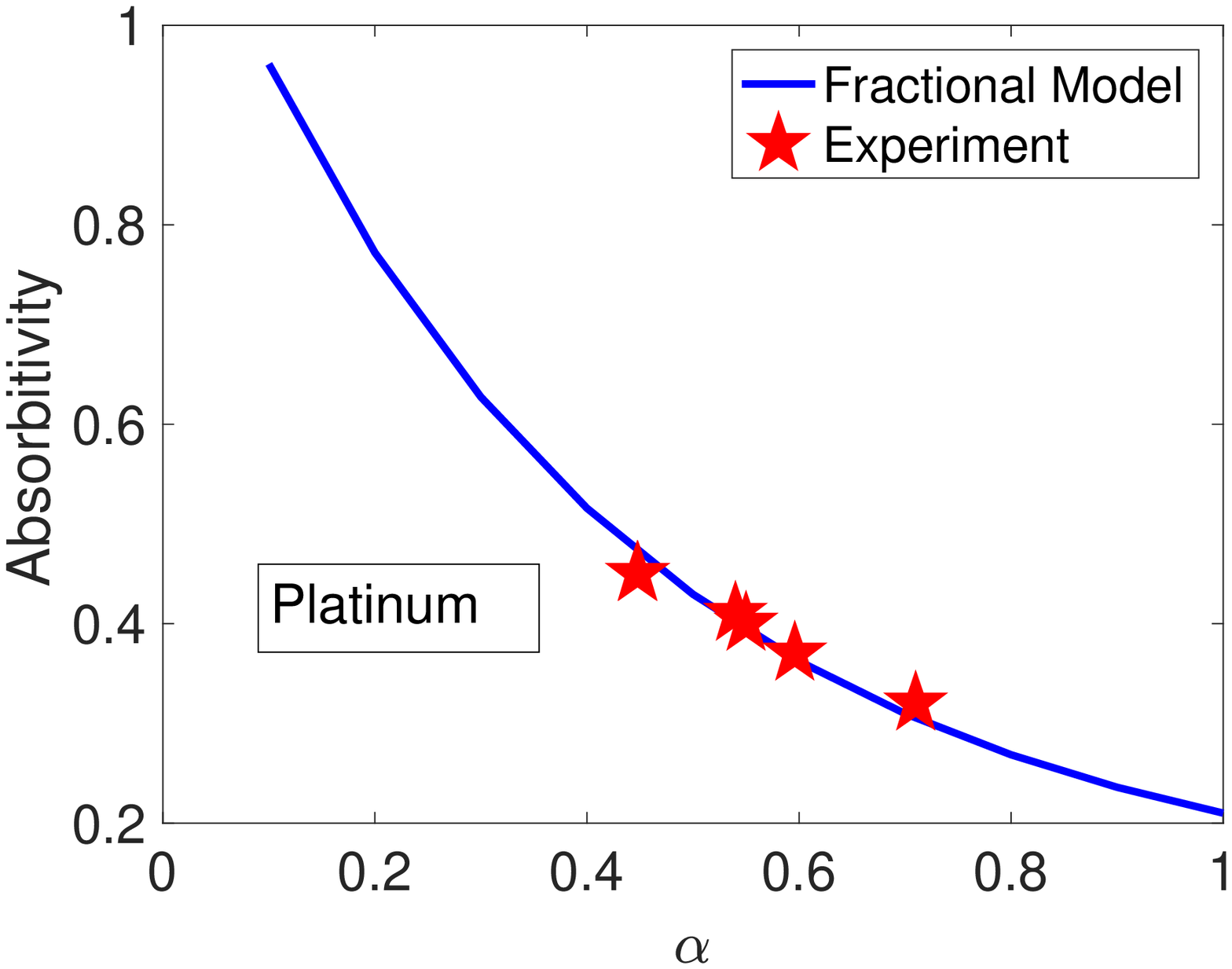}%
       		%\label{Fig:Pt}
       	} 
       	%                     		\subfloat[\label{fig:Zn_exp}]{%
       	%                     			\includegraphics[width=.22\textwidth]{Zn_exp}%
       	%                     			%\label{Fig:Pt}
       	%                     		} 
       	\caption{Comparison with experimentally measured absorption of (a) Cu~\cite{vorobyev2007change}, (b) Au~\cite{vorobyev2005enhanced}, and (c) Pt~\cite{vorobyev2007effects} surfaces as a function of fractal dimension parameter $\alpha$ determined through box-counting of rough surface SEM images.} \label{Fig:exp}
       \end{figure}

We applied same procedure to calculate $D$ from SEM images of \textcolor{red}{micro-scale rough surfaces} reported in Refs. \cite{vorobyev2007change,vorobyev2005enhanced,vorobyev2007effects}. 
The details can be found in the supplementary material. 
By using these values of $\alpha$ into Eq. (\ref{eqn:abs_coeeficient}), we analytically calculate the absorptivity for copper, gold and platinum as shown in Fig.~(\ref{Fig:exp}).
From the results, we see excellent agreements between the calculations (solid lines) and experimental measurements (symbols).
For each metal, the corresponding complex $\epsilon_{r}$ at $\lambda$=800 nm is taken from~\cite{werner2009optical}. 

\begin{figure}[!h]
        	\centering
        	\includegraphics[width=0.5\textwidth]{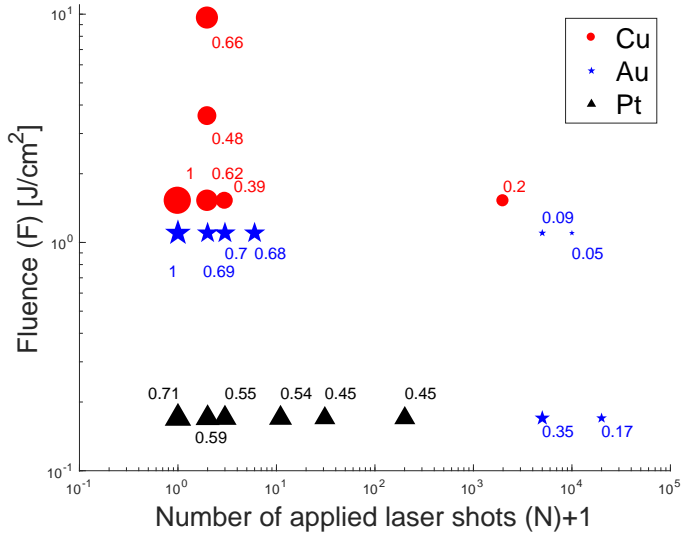}
        	\caption{Calculated $\alpha$ for rough surfaces of copper, gold and platinum at various combinations of laser fluence ($F$) and number of shots ($N$). Each point is labeled with the calculated value $\alpha$ and scaled in size accordingly. }
        	\label{fig:scatter_FNAlpha}
        \end{figure}

In Fig.~(\ref{fig:scatter_FNAlpha}), we show the estimated $\alpha$ for rough surfaces of copper \cite{vorobyev2007change}, gold \cite{vorobyev2005enhanced} and platinum~\cite{vorobyev2007effects} over a wide range of
laser fluence ($F$) and number of laser shots ($N$). 
It is observed that for platinum at small fluence $F$ (about 0.2 $\mathrm{J/cm^2}$) with an increasing number of laser shots ($N$ = 1 to 100), it is possible to have small $\alpha$ around 0.4 to 0.7 that may enhance the absorptivity to about 20 to 40 percent [see Fig. 4c].
For copper, at $N$ =1 by increasing $F$ from 1 to 10 $\mathrm{J/cm^2}$, absorptivity is also enhanced to more than 20 percent at $\alpha$ = 0.66. 
At fixed $F$ = 1 $\mathrm{J/cm^2}$, the roughness of copper can be increased to $\alpha$ = 0.39 with $N$ = 2 to 3.
At large $N>$ 5000 with $F$ = 1 $\mathrm{J/cm^2}$, gold can have a very rough surface with $\alpha$ = 0.05 to 0.09, which suggest a high absorptive gold with absorptivity more than 80 percent is possible. From this analysis, it is clear that the fractal dimension is a good measure in the optimization of laser processing conditions for designing of tunable absorptive metal surfaces.

%   we report ref\cite{vorobyev2007change}
%     we report from \cite{vorobyev2005enhanced}
%     we report from \cite{vorobyev2007effects, fang2017direct}

  \section{Summary} 
  
 The optical absorption of a rough metal surface is formulated by using a fractional model for which the absorptivity is analytically expressed as a function of $\alpha$ to account for the surface roughness.
By using the experimental results of fs laser induced roughness, we determine the values of $\alpha$ using box-counting method on reported SEM images of rough surfaces, and the calculated absorptivity shows very good agreements for 3 studied metals: gold, copper and platinum. This model provides a fast and useful approach for experiment to create suitable surface roughness in order to tune the optical absorptivity to a value needed for many applications that require highly absorptive metallic materials.
In terms of physics, this model can be seemed as a fractional model of the Fresnel coefficients.
Thus it is not limited to optical absorption reported here; where it can be used for the transmission and reflection of electromagnetic waves at any rough metallic surfaces or interfaces. It is worthwhile to mention that the box-counting method can show better results on high-precision atomic force microscopic (AFM) images. 
We emphasize the experimentalists to directly measure the optical absorption at various rough metal surfaces and record their respective AFM images of roughness profiles. The electronic availability of such optical absorption data along with high quality AFM images would be extremely useful for further explorations and improvements in the existing theoretical model.

         \section{Supplementary Material} 
        
        See supplementary material for the detailed fractal-dimension calculation of different rough metal (Cu, Au and Pt) surfaces using box-counting algorithm on SEM images reported in literature~\cite{vorobyev2007change,vorobyev2005enhanced,vorobyev2007effects}. 
        \begin{acknowledgements}
        	%We are very thankful to xxxx for helpful discussion. 
        	This work is supported by Singapore Temasek Laboratories (TL) seed grant (IGDS S16 02 05 1), and USA AFOSR AOARD grant (FA2386-14-1-4020).
        	
        \end{acknowledgements}
\clearpage

\bibliography{absorption_v3}

\providecommand{\noopsort}[1]{}\providecommand{\singleletter}[1]{#1}%
\begin{thebibliography}{43}%
\makeatletter
\providecommand \@ifxundefined [1]{%
 \@ifx{#1\undefined}
}%
\providecommand \@ifnum [1]{%
 \ifnum #1\expandafter \@firstoftwo
 \else \expandafter \@secondoftwo
 \fi
}%
\providecommand \@ifx [1]{%
 \ifx #1\expandafter \@firstoftwo
 \else \expandafter \@secondoftwo
 \fi
}%
\providecommand \natexlab [1]{#1}%
\providecommand \enquote  [1]{``#1''}%
\providecommand \bibnamefont  [1]{#1}%
\providecommand \bibfnamefont [1]{#1}%
\providecommand \citenamefont [1]{#1}%
\providecommand \href@noop [0]{\@secondoftwo}%
\providecommand \href [0]{\begingroup \@sanitize@url \@href}%
\providecommand \@href[1]{\@@startlink{#1}\@@href}%
\providecommand \@@href[1]{\endgroup#1\@@endlink}%
\providecommand \@sanitize@url [0]{\catcode `\\12\catcode `\$12\catcode
  `\&12\catcode `\#12\catcode `\^12\catcode `\_12\catcode `\%12\relax}%
\providecommand \@@startlink[1]{}%
\providecommand \@@endlink[0]{}%
\providecommand \url  [0]{\begingroup\@sanitize@url \@url }%
\providecommand \@url [1]{\endgroup\@href {#1}{\urlprefix }}%
\providecommand \urlprefix  [0]{URL }%
\providecommand \Eprint [0]{\href }%
\providecommand \doibase [0]{http://dx.doi.org/}%
\providecommand \selectlanguage [0]{\@gobble}%
\providecommand \bibinfo  [0]{\@secondoftwo}%
\providecommand \bibfield  [0]{\@secondoftwo}%
\providecommand \translation [1]{[#1]}%
\providecommand \BibitemOpen [0]{}%
\providecommand \bibitemStop [0]{}%
\providecommand \bibitemNoStop [0]{.\EOS\space}%
\providecommand \EOS [0]{\spacefactor3000\relax}%
\providecommand \BibitemShut  [1]{\csname bibitem#1\endcsname}%
\let\auto@bib@innerbib\@empty
%</preamble>
\bibitem [{\citenamefont {Gamaly}(2011)}]{gamaly2011femtosecond}%
  \BibitemOpen
  \bibfield  {author} {\bibinfo {author} {\bibfnamefont {E.~G.}\ \bibnamefont
  {Gamaly}},\ }\href@noop {} {\emph {\bibinfo {title} {Femtosecond Laser-Matter
  Interaction: Theory, Experiments and Applications}}}\ (\bibinfo  {publisher}
  {CRC Press},\ \bibinfo {year} {2011})\BibitemShut {NoStop}%
\bibitem [{\citenamefont {Vorobyev}\ and\ \citenamefont
  {Guo}(2013)}]{vorobyev2013direct}%
  \BibitemOpen
  \bibfield  {author} {\bibinfo {author} {\bibfnamefont {A.~Y.}\ \bibnamefont
  {Vorobyev}}\ and\ \bibinfo {author} {\bibfnamefont {C.}~\bibnamefont {Guo}},\
  }\href@noop {} {\bibfield  {journal} {\bibinfo  {journal} {Laser \& Photonics
  Reviews}\ }\textbf {\bibinfo {volume} {7}},\ \bibinfo {pages} {385} (\bibinfo
  {year} {2013})}\BibitemShut {NoStop}%
\bibitem [{\citenamefont {Vorobyev}\ and\ \citenamefont
  {Guo}(2005)}]{vorobyev2005enhanced}%
  \BibitemOpen
  \bibfield  {author} {\bibinfo {author} {\bibfnamefont {A.~Y.}\ \bibnamefont
  {Vorobyev}}\ and\ \bibinfo {author} {\bibfnamefont {C.}~\bibnamefont {Guo}},\
  }\href@noop {} {\bibfield  {journal} {\bibinfo  {journal} {Physical Review
  B}\ }\textbf {\bibinfo {volume} {72}},\ \bibinfo {pages} {195422} (\bibinfo
  {year} {2005})}\BibitemShut {NoStop}%
\bibitem [{\citenamefont {Vorobyev}\ \emph {et~al.}(2009)\citenamefont
  {Vorobyev}, \citenamefont {Topkov}, \citenamefont {Gurin}, \citenamefont
  {Svich},\ and\ \citenamefont {Guo}}]{vorobyev2009enhanced}%
  \BibitemOpen
  \bibfield  {author} {\bibinfo {author} {\bibfnamefont {A.~Y.}\ \bibnamefont
  {Vorobyev}}, \bibinfo {author} {\bibfnamefont {A.~N.}\ \bibnamefont
  {Topkov}}, \bibinfo {author} {\bibfnamefont {O.~V.}\ \bibnamefont {Gurin}},
  \bibinfo {author} {\bibfnamefont {V.~A.}\ \bibnamefont {Svich}}, \ and\
  \bibinfo {author} {\bibfnamefont {C.}~\bibnamefont {Guo}},\ }\href@noop {}
  {\bibfield  {journal} {\bibinfo  {journal} {Applied Physics Letters}\
  }\textbf {\bibinfo {volume} {95}},\ \bibinfo {pages} {121106} (\bibinfo
  {year} {2009})}\BibitemShut {NoStop}%
\bibitem [{\citenamefont {Fang}\ \emph {et~al.}(2017)\citenamefont {Fang},
  \citenamefont {Vorobyev},\ and\ \citenamefont {Guo}}]{fang2017direct}%
  \BibitemOpen
  \bibfield  {author} {\bibinfo {author} {\bibfnamefont {R.}~\bibnamefont
  {Fang}}, \bibinfo {author} {\bibfnamefont {A.}~\bibnamefont {Vorobyev}}, \
  and\ \bibinfo {author} {\bibfnamefont {C.}~\bibnamefont {Guo}},\ }\href@noop
  {} {\bibfield  {journal} {\bibinfo  {journal} {Light: Science and
  Applications}\ }\textbf {\bibinfo {volume} {6}} (\bibinfo {year}
  {2017})}\BibitemShut {NoStop}%
\bibitem [{\citenamefont {Sugioka}\ and\ \citenamefont
  {Cheng}(2014)}]{sugioka2014ultrafast}%
  \BibitemOpen
  \bibfield  {author} {\bibinfo {author} {\bibfnamefont {K.}~\bibnamefont
  {Sugioka}}\ and\ \bibinfo {author} {\bibfnamefont {Y.}~\bibnamefont
  {Cheng}},\ }\href@noop {} {\bibfield  {journal} {\bibinfo  {journal} {Light:
  Science and Applications}\ }\textbf {\bibinfo {volume} {3}},\ \bibinfo
  {pages} {e149} (\bibinfo {year} {2014})}\BibitemShut {NoStop}%
\bibitem [{\citenamefont {Ghmari}\ \emph {et~al.}(2004)\citenamefont {Ghmari},
  \citenamefont {Ghbara}, \citenamefont {Laroche}, \citenamefont {Carminati},\
  and\ \citenamefont {Greffet}}]{ghmari2004influence}%
  \BibitemOpen
  \bibfield  {author} {\bibinfo {author} {\bibfnamefont {F.}~\bibnamefont
  {Ghmari}}, \bibinfo {author} {\bibfnamefont {T.}~\bibnamefont {Ghbara}},
  \bibinfo {author} {\bibfnamefont {M.}~\bibnamefont {Laroche}}, \bibinfo
  {author} {\bibfnamefont {R.}~\bibnamefont {Carminati}}, \ and\ \bibinfo
  {author} {\bibfnamefont {J.-J.}\ \bibnamefont {Greffet}},\ }\href@noop {}
  {\bibfield  {journal} {\bibinfo  {journal} {Journal of Applied Physics}\
  }\textbf {\bibinfo {volume} {96}},\ \bibinfo {pages} {2656} (\bibinfo {year}
  {2004})}\BibitemShut {NoStop}%
\bibitem [{\citenamefont {Ang}\ \emph {et~al.}(1997)\citenamefont {Ang},
  \citenamefont {Lau}, \citenamefont {Gilgenbach},\ and\ \citenamefont
  {Spindler}}]{ang1997analysis}%
  \BibitemOpen
  \bibfield  {author} {\bibinfo {author} {\bibfnamefont {L.~K.}\ \bibnamefont
  {Ang}}, \bibinfo {author} {\bibfnamefont {Y.~Y.}\ \bibnamefont {Lau}},
  \bibinfo {author} {\bibfnamefont {R.~M.}\ \bibnamefont {Gilgenbach}}, \ and\
  \bibinfo {author} {\bibfnamefont {H.~L.}\ \bibnamefont {Spindler}},\
  }\href@noop {} {\bibfield  {journal} {\bibinfo  {journal} {Applied physics
  letters}\ }\textbf {\bibinfo {volume} {70}},\ \bibinfo {pages} {696}
  (\bibinfo {year} {1997})}\BibitemShut {NoStop}%
\bibitem [{\citenamefont {Gavrilenko}(2011)}]{gavrilenko2011optics}%
  \BibitemOpen
  \bibfield  {author} {\bibinfo {author} {\bibfnamefont {V.~I.}\ \bibnamefont
  {Gavrilenko}},\ }\href@noop {} {\emph {\bibinfo {title} {Optics of
  Nanomaterials}}}\ (\bibinfo  {publisher} {Pan Stanford Publishing},\ \bibinfo
  {year} {2011})\BibitemShut {NoStop}%
\bibitem [{\citenamefont {Bergstr{\"o}m}(2008)}]{bergstrom2008absorption}%
  \BibitemOpen
  \bibfield  {author} {\bibinfo {author} {\bibfnamefont {D.}~\bibnamefont
  {Bergstr{\"o}m}},\ }\emph {\bibinfo {title} {The absorption of laser light by
  rough metal surfaces}},\ \href@noop {} {Ph.D. thesis},\ \bibinfo  {school}
  {Lule{\aa} tekniska universitet} (\bibinfo {year} {2008})\BibitemShut
  {NoStop}%
\bibitem [{\citenamefont {Mandelbrot}\ and\ \citenamefont
  {Pignoni}(1983)}]{mandelbrot1983fractal}%
  \BibitemOpen
  \bibfield  {author} {\bibinfo {author} {\bibfnamefont {B.~B.}\ \bibnamefont
  {Mandelbrot}}\ and\ \bibinfo {author} {\bibfnamefont {R.}~\bibnamefont
  {Pignoni}},\ }\href@noop {} {\emph {\bibinfo {title} {The fractal geometry of
  nature}}},\ Vol.\ \bibinfo {volume} {173}\ (\bibinfo  {publisher} {WH freeman
  New York},\ \bibinfo {year} {1983})\BibitemShut {NoStop}%
\bibitem [{\citenamefont {Falconer}(2004)}]{falconer2004fractal}%
  \BibitemOpen
  \bibfield  {author} {\bibinfo {author} {\bibfnamefont {K.}~\bibnamefont
  {Falconer}},\ }\href@noop {} {\emph {\bibinfo {title} {Fractal geometry:
  mathematical foundations and applications}}}\ (\bibinfo  {publisher} {John
  Wiley \& Sons},\ \bibinfo {year} {2004})\BibitemShut {NoStop}%
\bibitem [{\citenamefont {Risovi{\'c}}\ \emph {et~al.}(2009)\citenamefont
  {Risovi{\'c}}, \citenamefont {Polja{\v{c}}ek},\ and\ \citenamefont
  {Gojo}}]{risovic2009correlation}%
  \BibitemOpen
  \bibfield  {author} {\bibinfo {author} {\bibfnamefont {D.}~\bibnamefont
  {Risovi{\'c}}}, \bibinfo {author} {\bibfnamefont {S.~M.}\ \bibnamefont
  {Polja{\v{c}}ek}}, \ and\ \bibinfo {author} {\bibfnamefont {M.}~\bibnamefont
  {Gojo}},\ }\href@noop {} {\bibfield  {journal} {\bibinfo  {journal} {Applied
  Surface Science}\ }\textbf {\bibinfo {volume} {255}},\ \bibinfo {pages}
  {4283} (\bibinfo {year} {2009})}\BibitemShut {NoStop}%
\bibitem [{\citenamefont {Risovi{\'c}}\ \emph {et~al.}(2008)\citenamefont
  {Risovi{\'c}}, \citenamefont {Polja{\v{c}}ek}, \citenamefont {Furi{\'c}},\
  and\ \citenamefont {Gojo}}]{risovic2008inferring}%
  \BibitemOpen
  \bibfield  {author} {\bibinfo {author} {\bibfnamefont {D.}~\bibnamefont
  {Risovi{\'c}}}, \bibinfo {author} {\bibfnamefont {S.~M.}\ \bibnamefont
  {Polja{\v{c}}ek}}, \bibinfo {author} {\bibfnamefont {K.}~\bibnamefont
  {Furi{\'c}}}, \ and\ \bibinfo {author} {\bibfnamefont {M.}~\bibnamefont
  {Gojo}},\ }\href@noop {} {\bibfield  {journal} {\bibinfo  {journal} {Applied
  Surface Science}\ }\textbf {\bibinfo {volume} {255}},\ \bibinfo {pages}
  {3063} (\bibinfo {year} {2008})}\BibitemShut {NoStop}%
\bibitem [{\citenamefont {Russ}(2001)}]{russ2001fractal}%
  \BibitemOpen
  \bibfield  {author} {\bibinfo {author} {\bibfnamefont {J.~C.}\ \bibnamefont
  {Russ}},\ }in\ \href@noop {} {\emph {\bibinfo {booktitle} {Metrology and
  Properties of Engineering Surfaces}}}\ (\bibinfo  {publisher} {Springer},\
  \bibinfo {year} {2001})\ pp.\ \bibinfo {pages} {43--82}\BibitemShut {NoStop}%
\bibitem [{\citenamefont {Vorobyev}\ and\ \citenamefont
  {Guo}(2007{\natexlab{a}})}]{vorobyev2007change}%
  \BibitemOpen
  \bibfield  {author} {\bibinfo {author} {\bibfnamefont {A.~Y.}\ \bibnamefont
  {Vorobyev}}\ and\ \bibinfo {author} {\bibfnamefont {C.}~\bibnamefont {Guo}},\
  }in\ \href@noop {} {\emph {\bibinfo {booktitle} {Journal of Physics:
  Conference Series}}},\ Vol.~\bibinfo {volume} {59}\ (\bibinfo {organization}
  {IOP Publishing},\ \bibinfo {year} {2007})\ p.\ \bibinfo {pages}
  {579}\BibitemShut {NoStop}%
\bibitem [{\citenamefont {Vorobyev}\ and\ \citenamefont
  {Guo}(2007{\natexlab{b}})}]{vorobyev2007effects}%
  \BibitemOpen
  \bibfield  {author} {\bibinfo {author} {\bibfnamefont {A.~Y.}\ \bibnamefont
  {Vorobyev}}\ and\ \bibinfo {author} {\bibfnamefont {C.}~\bibnamefont {Guo}},\
  }\href@noop {} {\bibfield  {journal} {\bibinfo  {journal} {Applied Physics A:
  Materials Science \& Processing}\ }\textbf {\bibinfo {volume} {86}},\
  \bibinfo {pages} {321} (\bibinfo {year} {2007}{\natexlab{b}})}\BibitemShut
  {NoStop}%
\bibitem [{\citenamefont {Zubair}\ \emph {et~al.}(2012)\citenamefont {Zubair},
  \citenamefont {Mughal},\ and\ \citenamefont
  {Naqvi}}]{zubair2012electromagnetic}%
  \BibitemOpen
  \bibfield  {author} {\bibinfo {author} {\bibfnamefont {M.}~\bibnamefont
  {Zubair}}, \bibinfo {author} {\bibfnamefont {M.~J.}\ \bibnamefont {Mughal}},
  \ and\ \bibinfo {author} {\bibfnamefont {Q.~A.}\ \bibnamefont {Naqvi}},\
  }\href@noop {} {\emph {\bibinfo {title} {Electromagnetic fields and waves in
  fractional dimensional space}}}\ (\bibinfo  {publisher} {Springer Science \&
  Business Media},\ \bibinfo {year} {2012})\BibitemShut {NoStop}%
\bibitem [{\citenamefont {Tarasov}(2014)}]{tarasov2014anisotropic}%
  \BibitemOpen
  \bibfield  {author} {\bibinfo {author} {\bibfnamefont {V.~E.}\ \bibnamefont
  {Tarasov}},\ }\href@noop {} {\bibfield  {journal} {\bibinfo  {journal}
  {Journal of Mathematical Physics}\ }\textbf {\bibinfo {volume} {55}},\
  \bibinfo {pages} {083510} (\bibinfo {year} {2014})}\BibitemShut {NoStop}%
\bibitem [{\citenamefont {Stillinger}(1977)}]{stillinger1977axiomatic}%
  \BibitemOpen
  \bibfield  {author} {\bibinfo {author} {\bibfnamefont {F.~H.}\ \bibnamefont
  {Stillinger}},\ }\href@noop {} {\bibfield  {journal} {\bibinfo  {journal}
  {Journal of Mathematical Physics}\ }\textbf {\bibinfo {volume} {18}},\
  \bibinfo {pages} {1224} (\bibinfo {year} {1977})}\BibitemShut {NoStop}%
\bibitem [{\citenamefont {Palmer}\ and\ \citenamefont
  {Stavrinou}(2004)}]{palmer2004equations}%
  \BibitemOpen
  \bibfield  {author} {\bibinfo {author} {\bibfnamefont {C.}~\bibnamefont
  {Palmer}}\ and\ \bibinfo {author} {\bibfnamefont {P.~N.}\ \bibnamefont
  {Stavrinou}},\ }\href@noop {} {\bibfield  {journal} {\bibinfo  {journal}
  {Journal of Physics A: Mathematical and General}\ }\textbf {\bibinfo {volume}
  {37}},\ \bibinfo {pages} {6987} (\bibinfo {year} {2004})}\BibitemShut
  {NoStop}%
\bibitem [{\citenamefont {Sadallah}\ and\ \citenamefont
  {Muslih}(2009)}]{sadallah2009solution}%
  \BibitemOpen
  \bibfield  {author} {\bibinfo {author} {\bibfnamefont {M.}~\bibnamefont
  {Sadallah}}\ and\ \bibinfo {author} {\bibfnamefont {S.~I.}\ \bibnamefont
  {Muslih}},\ }\href@noop {} {\bibfield  {journal} {\bibinfo  {journal}
  {International Journal of Theoretical Physics}\ }\textbf {\bibinfo {volume}
  {48}},\ \bibinfo {pages} {3312} (\bibinfo {year} {2009})}\BibitemShut
  {NoStop}%
\bibitem [{\citenamefont {Tarasov}(2016)}]{tarasov2016heat}%
  \BibitemOpen
  \bibfield  {author} {\bibinfo {author} {\bibfnamefont {V.~E.}\ \bibnamefont
  {Tarasov}},\ }\href@noop {} {\bibfield  {journal} {\bibinfo  {journal}
  {International Journal of Heat and Mass Transfer}\ }\textbf {\bibinfo
  {volume} {93}},\ \bibinfo {pages} {427} (\bibinfo {year} {2016})}\BibitemShut
  {NoStop}%
\bibitem [{\citenamefont {Ostoja-Starzewski}\ \emph {et~al.}(2014)\citenamefont
  {Ostoja-Starzewski}, \citenamefont {Li}, \citenamefont {Joumaa},\ and\
  \citenamefont {Demmie}}]{ostoja2014fractal}%
  \BibitemOpen
  \bibfield  {author} {\bibinfo {author} {\bibfnamefont {M.}~\bibnamefont
  {Ostoja-Starzewski}}, \bibinfo {author} {\bibfnamefont {J.}~\bibnamefont
  {Li}}, \bibinfo {author} {\bibfnamefont {H.}~\bibnamefont {Joumaa}}, \ and\
  \bibinfo {author} {\bibfnamefont {P.~N.}\ \bibnamefont {Demmie}},\
  }\href@noop {} {\bibfield  {journal} {\bibinfo  {journal} {ZAMM-Journal of
  Applied Mathematics and Mechanics/Zeitschrift f{\"u}r Angewandte Mathematik
  und Mechanik}\ }\textbf {\bibinfo {volume} {94}},\ \bibinfo {pages} {373}
  (\bibinfo {year} {2014})}\BibitemShut {NoStop}%
\bibitem [{\citenamefont {Balankin}\ and\ \citenamefont
  {Elizarraraz}(2012)}]{balankin2012map}%
  \BibitemOpen
  \bibfield  {author} {\bibinfo {author} {\bibfnamefont {A.~S.}\ \bibnamefont
  {Balankin}}\ and\ \bibinfo {author} {\bibfnamefont {B.~E.}\ \bibnamefont
  {Elizarraraz}},\ }\href@noop {} {\bibfield  {journal} {\bibinfo  {journal}
  {Physical Review E}\ }\textbf {\bibinfo {volume} {85}},\ \bibinfo {pages}
  {056314} (\bibinfo {year} {2012})}\BibitemShut {NoStop}%
\bibitem [{\citenamefont {Tarasov}(2015)}]{tarasov2015electromagnetic}%
  \BibitemOpen
  \bibfield  {author} {\bibinfo {author} {\bibfnamefont {V.~E.}\ \bibnamefont
  {Tarasov}},\ }\href@noop {} {\bibfield  {journal} {\bibinfo  {journal}
  {Chaos, Solitons \& Fractals}\ }\textbf {\bibinfo {volume} {81}},\ \bibinfo
  {pages} {38} (\bibinfo {year} {2015})}\BibitemShut {NoStop}%
\bibitem [{\citenamefont {Mughal}\ and\ \citenamefont
  {Zubair}(2011)}]{mughal2011fractional}%
  \BibitemOpen
  \bibfield  {author} {\bibinfo {author} {\bibfnamefont {M.~J.}\ \bibnamefont
  {Mughal}}\ and\ \bibinfo {author} {\bibfnamefont {M.}~\bibnamefont
  {Zubair}},\ }in\ \href@noop {} {\emph {\bibinfo {booktitle} {Signal
  Processing and Communications Applications (SIU), 2011 IEEE 19th Conference
  on}}}\ (\bibinfo {organization} {IEEE},\ \bibinfo {year} {2011})\ pp.\
  \bibinfo {pages} {62--65}\BibitemShut {NoStop}%
\bibitem [{\citenamefont {Naqvi}\ and\ \citenamefont
  {Zubair}(2016)}]{naqvi2016cylindrical}%
  \BibitemOpen
  \bibfield  {author} {\bibinfo {author} {\bibfnamefont {Q.~A.}\ \bibnamefont
  {Naqvi}}\ and\ \bibinfo {author} {\bibfnamefont {M.}~\bibnamefont {Zubair}},\
  }\href@noop {} {\bibfield  {journal} {\bibinfo  {journal}
  {Optik-International Journal for Light and Electron Optics}\ }\textbf
  {\bibinfo {volume} {127}},\ \bibinfo {pages} {3243} (\bibinfo {year}
  {2016})}\BibitemShut {NoStop}%
\bibitem [{\citenamefont {Zubair}\ \emph
  {et~al.}(2011{\natexlab{a}})\citenamefont {Zubair}, \citenamefont {Mughal},\
  and\ \citenamefont {Naqvi}}]{zubair2011exact}%
  \BibitemOpen
  \bibfield  {author} {\bibinfo {author} {\bibfnamefont {M.}~\bibnamefont
  {Zubair}}, \bibinfo {author} {\bibfnamefont {M.~J.}\ \bibnamefont {Mughal}},
  \ and\ \bibinfo {author} {\bibfnamefont {Q.~A.}\ \bibnamefont {Naqvi}},\
  }\href@noop {} {\bibfield  {journal} {\bibinfo  {journal} {Progress In
  Electromagnetics Research}\ }\textbf {\bibinfo {volume} {114}},\ \bibinfo
  {pages} {443} (\bibinfo {year} {2011}{\natexlab{a}})}\BibitemShut {NoStop}%
\bibitem [{\citenamefont {Asad}\ \emph
  {et~al.}(2012{\natexlab{a}})\citenamefont {Asad}, \citenamefont {Mughal},
  \citenamefont {Zubair},\ and\ \citenamefont
  {Naqvi}}]{asad2012electromagnetic}%
  \BibitemOpen
  \bibfield  {author} {\bibinfo {author} {\bibfnamefont {H.}~\bibnamefont
  {Asad}}, \bibinfo {author} {\bibfnamefont {M.~J.}\ \bibnamefont {Mughal}},
  \bibinfo {author} {\bibfnamefont {M.}~\bibnamefont {Zubair}}, \ and\ \bibinfo
  {author} {\bibfnamefont {Q.~A.}\ \bibnamefont {Naqvi}},\ }\href@noop {}
  {\bibfield  {journal} {\bibinfo  {journal} {Journal of Electromagnetic Waves
  and Applications}\ }\textbf {\bibinfo {volume} {26}},\ \bibinfo {pages}
  {1903} (\bibinfo {year} {2012}{\natexlab{a}})}\BibitemShut {NoStop}%
\bibitem [{\citenamefont {Asad}\ \emph
  {et~al.}(2012{\natexlab{b}})\citenamefont {Asad}, \citenamefont {Zubair},\
  and\ \citenamefont {Mughal}}]{asad2012reflection}%
  \BibitemOpen
  \bibfield  {author} {\bibinfo {author} {\bibfnamefont {H.}~\bibnamefont
  {Asad}}, \bibinfo {author} {\bibfnamefont {M.}~\bibnamefont {Zubair}}, \ and\
  \bibinfo {author} {\bibfnamefont {M.~J.}\ \bibnamefont {Mughal}},\
  }\href@noop {} {\bibfield  {journal} {\bibinfo  {journal} {Progress In
  Electromagnetics Research}\ }\textbf {\bibinfo {volume} {125}},\ \bibinfo
  {pages} {543} (\bibinfo {year} {2012}{\natexlab{b}})}\BibitemShut {NoStop}%
\bibitem [{\citenamefont {Zubair}\ \emph
  {et~al.}(2011{\natexlab{b}})\citenamefont {Zubair}, \citenamefont {Mughal},\
  and\ \citenamefont {Naqvi}}]{zubair2011exact2}%
  \BibitemOpen
  \bibfield  {author} {\bibinfo {author} {\bibfnamefont {M.}~\bibnamefont
  {Zubair}}, \bibinfo {author} {\bibfnamefont {M.~J.}\ \bibnamefont {Mughal}},
  \ and\ \bibinfo {author} {\bibfnamefont {Q.~A.}\ \bibnamefont {Naqvi}},\
  }\href@noop {} {\bibfield  {journal} {\bibinfo  {journal} {Journal of
  Electromagnetic Waves and Applications}\ }\textbf {\bibinfo {volume} {25}},\
  \bibinfo {pages} {1481} (\bibinfo {year} {2011}{\natexlab{b}})}\BibitemShut
  {NoStop}%
\bibitem [{\citenamefont {Zubair}\ \emph
  {et~al.}(2011{\natexlab{c}})\citenamefont {Zubair}, \citenamefont {Mughal},
  \citenamefont {Naqvi},\ and\ \citenamefont {Rizvi}}]{zubair2011differential}%
  \BibitemOpen
  \bibfield  {author} {\bibinfo {author} {\bibfnamefont {M.}~\bibnamefont
  {Zubair}}, \bibinfo {author} {\bibfnamefont {M.~J.}\ \bibnamefont {Mughal}},
  \bibinfo {author} {\bibfnamefont {Q.~A.}\ \bibnamefont {Naqvi}}, \ and\
  \bibinfo {author} {\bibfnamefont {A.~A.}\ \bibnamefont {Rizvi}},\ }\href@noop
  {} {\bibfield  {journal} {\bibinfo  {journal} {Progress In Electromagnetics
  Research}\ }\textbf {\bibinfo {volume} {114}},\ \bibinfo {pages} {255}
  (\bibinfo {year} {2011}{\natexlab{c}})}\BibitemShut {NoStop}%
\bibitem [{\citenamefont {Zubair}\ \emph
  {et~al.}(2011{\natexlab{d}})\citenamefont {Zubair}, \citenamefont {Mughal},\
  and\ \citenamefont {Naqvi}}]{zubair2011electromagnetic}%
  \BibitemOpen
  \bibfield  {author} {\bibinfo {author} {\bibfnamefont {M.}~\bibnamefont
  {Zubair}}, \bibinfo {author} {\bibfnamefont {M.~J.}\ \bibnamefont {Mughal}},
  \ and\ \bibinfo {author} {\bibfnamefont {Q.~A.}\ \bibnamefont {Naqvi}},\
  }\href@noop {} {\bibfield  {journal} {\bibinfo  {journal} {Nonlinear
  Analysis: Real World Applications}\ }\textbf {\bibinfo {volume} {12}},\
  \bibinfo {pages} {2844} (\bibinfo {year} {2011}{\natexlab{d}})}\BibitemShut
  {NoStop}%
\bibitem [{\citenamefont {\vspace{0mm}Zubair}\ \emph
  {et~al.}(2010)\citenamefont {\vspace{0mm}Zubair}, \citenamefont {Mughal},\
  and\ \citenamefont {Naqvi}}]{zubair2010wave}%
  \BibitemOpen
  \bibfield  {author} {\bibinfo {author} {\bibfnamefont {M.}~\bibnamefont
  {\vspace{0mm}Zubair}}, \bibinfo {author} {\bibfnamefont {M.~J.}\ \bibnamefont
  {Mughal}}, \ and\ \bibinfo {author} {\bibfnamefont {Q.~A.}\ \bibnamefont
  {Naqvi}},\ }\href@noop {} {\bibfield  {journal} {\bibinfo  {journal}
  {Progress In Electromagnetics Research Letters}\ }\textbf {\bibinfo {volume}
  {19}},\ \bibinfo {pages} {137} (\bibinfo {year} {2010})}\BibitemShut
  {NoStop}%
\bibitem [{\citenamefont {Zubair}\ and\ \citenamefont
  {Ang}(2016)}]{zubair2016fractional}%
  \BibitemOpen
  \bibfield  {author} {\bibinfo {author} {\bibfnamefont {M.}~\bibnamefont
  {Zubair}}\ and\ \bibinfo {author} {\bibfnamefont {L.~K.}\ \bibnamefont
  {Ang}},\ }\href@noop {} {\bibfield  {journal} {\bibinfo  {journal} {Physics
  of Plasmas (1994-present)}\ }\textbf {\bibinfo {volume} {23}},\ \bibinfo
  {pages} {072118} (\bibinfo {year} {2016})}\BibitemShut {NoStop}%
\bibitem [{\citenamefont {Zubair}\ \emph
  {et~al.}(2018{\natexlab{a}})\citenamefont {Zubair}, \citenamefont {Ang},\
  and\ \citenamefont {Ang}}]{zubair2017fractional}%
  \BibitemOpen
  \bibfield  {author} {\bibinfo {author} {\bibfnamefont {M.}~\bibnamefont
  {Zubair}}, \bibinfo {author} {\bibfnamefont {Y.~S.}\ \bibnamefont {Ang}}, \
  and\ \bibinfo {author} {\bibfnamefont {L.~K.}\ \bibnamefont {Ang}},\
  }\href@noop {} {\bibfield  {journal} {\bibinfo  {journal} {IEEE Transactions
  on Electron Devices}\ }\textbf {\bibinfo {volume} {65}},\ \bibinfo {pages}
  {2089} (\bibinfo {year} {2018}{\natexlab{a}})}\BibitemShut {NoStop}%
\bibitem [{\citenamefont {Zubair}\ \emph
  {et~al.}(2018{\natexlab{b}})\citenamefont {Zubair}, \citenamefont {Ang},\
  and\ \citenamefont {Ang}}]{zubair2018thickness}%
  \BibitemOpen
  \bibfield  {author} {\bibinfo {author} {\bibfnamefont {M.}~\bibnamefont
  {Zubair}}, \bibinfo {author} {\bibfnamefont {Y.~S.}\ \bibnamefont {Ang}}, \
  and\ \bibinfo {author} {\bibfnamefont {L.~K.}\ \bibnamefont {Ang}},\ }\href
  {\doibase 10.1109/TED.2018.2841920} {\bibfield  {journal} {\bibinfo
  {journal} {IEEE Transactions on Electron Devices, DOI:
  10.1109/TED.2018.2841920}\ ,\ \bibinfo {pages} {in}} (\bibinfo {year}
  {2018}{\natexlab{b}})}\BibitemShut {NoStop}%
\bibitem [{\citenamefont {Born}\ and\ \citenamefont
  {Wolf}(2013)}]{born2013principles}%
  \BibitemOpen
  \bibfield  {author} {\bibinfo {author} {\bibfnamefont {M.}~\bibnamefont
  {Born}}\ and\ \bibinfo {author} {\bibfnamefont {E.}~\bibnamefont {Wolf}},\
  }\href@noop {} {\emph {\bibinfo {title} {Principles of optics:
  electromagnetic theory of propagation, interference and diffraction of
  light}}}\ (\bibinfo  {publisher} {Elsevier},\ \bibinfo {year}
  {2013})\BibitemShut {NoStop}%
\bibitem [{\citenamefont {Moisy}(2008)}]{moisy2008computing}%
  \BibitemOpen
  \bibfield  {author} {\bibinfo {author} {\bibfnamefont {F.}~\bibnamefont
  {Moisy}},\ }\href@noop {} {\bibfield  {journal} {\bibinfo  {journal}
  {Laboratory FAST, University Paris Sud. Paris
  (http://www.fast.u-psud.fr/~moisy/ml/boxcount/html/demo.html)}\ } (\bibinfo
  {year} {2008})}\BibitemShut {NoStop}%
\bibitem [{\citenamefont {Davis}\ \emph {et~al.}(2017)\citenamefont {Davis},
  \citenamefont {Liu}, \citenamefont {Jiang}, \citenamefont {Lu},\ and\
  \citenamefont {Ndao}}]{davis2017wetting}%
  \BibitemOpen
  \bibfield  {author} {\bibinfo {author} {\bibfnamefont {E.}~\bibnamefont
  {Davis}}, \bibinfo {author} {\bibfnamefont {Y.}~\bibnamefont {Liu}}, \bibinfo
  {author} {\bibfnamefont {L.}~\bibnamefont {Jiang}}, \bibinfo {author}
  {\bibfnamefont {Y.}~\bibnamefont {Lu}}, \ and\ \bibinfo {author}
  {\bibfnamefont {S.}~\bibnamefont {Ndao}},\ }\href@noop {} {\bibfield
  {journal} {\bibinfo  {journal} {Applied Surface Science}\ }\textbf {\bibinfo
  {volume} {392}},\ \bibinfo {pages} {929} (\bibinfo {year}
  {2017})}\BibitemShut {NoStop}%
\bibitem [{\citenamefont {Chen}\ \emph {et~al.}(2017)\citenamefont {Chen},
  \citenamefont {Ma}, \citenamefont {Wang}, \citenamefont {He}, \citenamefont
  {Liu}, \citenamefont {Wang},\ and\ \citenamefont {Xu}}]{chen2017comparison}%
  \BibitemOpen
  \bibfield  {author} {\bibinfo {author} {\bibfnamefont {S.-y.}\ \bibnamefont
  {Chen}}, \bibinfo {author} {\bibfnamefont {G.-z.}\ \bibnamefont {Ma}},
  \bibinfo {author} {\bibfnamefont {H.-d.}\ \bibnamefont {Wang}}, \bibinfo
  {author} {\bibfnamefont {P.-f.}\ \bibnamefont {He}}, \bibinfo {author}
  {\bibfnamefont {M.}~\bibnamefont {Liu}}, \bibinfo {author} {\bibfnamefont
  {H.-j.}\ \bibnamefont {Wang}}, \ and\ \bibinfo {author} {\bibfnamefont
  {B.-s.}\ \bibnamefont {Xu}},\ }\href@noop {} {\bibfield  {journal} {\bibinfo
  {journal} {Applied Surface Science}\ }\textbf {\bibinfo {volume} {409}},\
  \bibinfo {pages} {277} (\bibinfo {year} {2017})}\BibitemShut {NoStop}%
\bibitem [{\citenamefont {Werner}\ \emph {et~al.}(2009)\citenamefont {Werner},
  \citenamefont {Glantschnig},\ and\ \citenamefont
  {Ambrosch-Draxl}}]{werner2009optical}%
  \BibitemOpen
  \bibfield  {author} {\bibinfo {author} {\bibfnamefont {W.~S.~M.}\
  \bibnamefont {Werner}}, \bibinfo {author} {\bibfnamefont {K.}~\bibnamefont
  {Glantschnig}}, \ and\ \bibinfo {author} {\bibfnamefont {C.}~\bibnamefont
  {Ambrosch-Draxl}},\ }\href@noop {} {\bibfield  {journal} {\bibinfo  {journal}
  {Journal of Physical and Chemical Reference Data}\ }\textbf {\bibinfo
  {volume} {38}},\ \bibinfo {pages} {1013} (\bibinfo {year}
  {2009})}\BibitemShut {NoStop}%
\end{thebibliography}%
\end{document}